\documentclass{emulateapj}

\usepackage{hyperref}

\begin{document}


\title{What a Tangled Web We Weave:  Hermus as the Northern Extension of the Phoenix Stream}

\author{Carl J. Grillmair}
\affil{Infrared Processing and Analysis Center, California Institute of Technology, Pasadena, CA 91125}
\email{carl@ipac.caltech.edu}

\author{Raymond G. Carlberg}
\affil{Department of Astronomy and Astrophysics, University of Toronto, Toronto, ON M5S 3H4, Canada}
\email{carlberg@astro.utoronto.ca}

\begin{abstract}

We investigate whether the recently discovered Phoenix stream may be
part of a much longer structure that includes the previously discovered
Hermus stream. Using a simple model of the Galaxy with a disk, bulge,
and a spherical dark matter halo, we show that a nearly circular
orbit, highly inclined with respect to the disk, can be found that fits the
positions, orientations, and distances of both streams. While the two
streams are somewhat misaligned in the sense that they do not occupy
the same plane, nodal precession due to the Milky Way disk potential
naturally brings the orbit into line with each stream in the course of
half an orbit.  We consequently consider a common origin for the two
streams as plausible. Based on our best fitting orbit, we make
predictions for the positions, distances, radial velocities, and proper
motions along each stream. If our hypothesis is borne out by
measurements, then at $\approx 183\arcdeg$ ($\approx 235\arcdeg$ with
respect to the Galactic center) and $\approx 76$ kpc in length,
Phoenix-Hermus would become the longest cold stream yet found. This
would make it a particularly valuable new probe of the shape and mass
of the Galactic halo out to $\approx 20$ kpc.

\end{abstract}


\keywords{Galaxy: Structure --- Galaxy: Halo}

\section{Introduction}

Recent years have seen the discovery of dozens of stellar debris
streams in our Galaxy (see \citet{grillmair2016} and \citet{smith2016}
for reviews). By virtue of their very low velocity dispersions, the
cold stellar streams we believe to be the remnants of globular
clusters are particularly well suited to the task of constraining the
shape and size of the Galactic potential. On the other hand, such
streams are far less populous than dwarf galaxy streams such as 
Orphan or Sagittarius and are consequently much harder to detect. Perhaps
for this reason, we have yet to find a cold stream that even
approaches the length of the Sagittarius stream.

Spanning $\approx 70\arcdeg$, GD-1 is the longest of the cold streams
discovered to date \citep{grillmair2006, carlberg2013}, though at a
distance of $\simeq 10$ kpc this corresponds to only 12 kpc in
length. Fitting orbits, \citet{koposov2010} used GD-1 to put
significant constraints on the circular velocity at the Sun's radius,
though they were rather insensitive to halo flattening due to the
proximity of the disk. \citet{eyre2011} noted that orbit-fitting is
not generally appropriate for stellar streams in realistic potentials,
as stars in streams do not follow a single orbit. Techniques have
recently been developed to quickly generate mock tidal streams without
a full N-body treatment and to match these steams directly to
observations \citep{lane2012,kupper2012,fardal2015}. Using such
techniques \citet{kupper2015} have once again demonstrated the potential of
cold streams as true ``high-precision scales'' of the Galactic mass
distribution.

Our expectation is that globular cluster streams should generally be
very long. Since tidal stripping will have begun virtually as soon as
globular clusters were born, then in the absence of major mergers \citep{wyse2009}, many streams will have had nearly the age of
the Universe to grow in length. Perturbations by dwarf galaxies, dark
matter subhalos, or disk structures can generate significant gaps in
streams \citep{carlberg2009, yoon2011}, or even decollimate large parts of
them to the point of undetectability. However, the existence of GD-1
demonstrates that such events are not common enough to shorten or
destroy {\it all} streams, at least in the inner halo.

From the work of \citet{kupper2015} and others we know that the
constraints that can be put on the shape and size of the Galactic
potential depend strongly on the length of a stream. We would clearly
benefit from the discovery of cold streams that extend completely
around the Galaxy, and at large distances from the Galactic center.

In this letter we test the hypothesis that the recently discovered
Phoenix and Hermus streams may be part of the same structure. We use a
simple model of the Galaxy to fit the streams in Section
\ref{analysis}. Based on our best fit to the streams, we make
predictions on observable quantities in Section \ref{predictions}.
Concluding remarks are given in Section \ref{conclusions}.

\section{Analysis} \label{analysis}

\subsection{Phoenix}

Balbinot (2015, hereafter B15) recently discovered a cold stream in
the Dark Energy Survey Year 1 data that they dubbed the Phoenix
stream. With a width of only 54 pc, this 17.5 kpc-distant,
$8\arcdeg$-long stream is very similar to the Pal 5 stream and
presumably also originated in a globular cluster. From the
color-magnitude distribution of its stars, B15 estimate an age of
11.5 Gyr and a metallicity of [Fe/H] $< -1.6$. Among the several
overdensities along the stream, B15 find a pair of peaks somewhat out
of alignment with the rest of the stream that they suggest may be
the remnant of the progenitor. If true, then the positions of these
peaks suggest that the stream is moving from south to north, in a
prograde orbit around the Galaxy.

\subsection{Hermus}

Grillmair (2014, hereafter G14) discovered a pair of nearly parallel
streams in the northern footprint of data release 10 of the Sloan
Digital Sky Survey (SDSS) \citep{ahn2014} that he dubbed Hermus and
Hyllus. Both streams appear to be metal poor, though the
color-magnitude distributions are very noisy and G14 was unable to
rule out [Fe/H] as high as -1.2 for Hermus. G14 estimated Hermus to be
about 20 kpc distant, with the northernmost end of the stream perhaps
as close as 15 kpc. While the stream appears to be some $50\arcdeg$
long (limited on both ends by limits of the SDSS survey footprint),
G14 noted that the southern $20\arcdeg$ had a somewhat different
character and curvature than the northern $30\arcdeg$, angling back
towards the west and becoming somewhat stronger and broader at the
southern end.

\section{Phoenix-Hermus?}

To search for possible Phoenix progenitors, B15 fit a great circle
that contained both the Phoenix stream and the Galactic center. They
found no globular clusters with similar distances along this great
circle and concluded that the stream could not have originated from
among the known clusters. In Panel a of Figure 1 we show this great
circle overplotted on the portion of the SDSS footprint containing
Hermus. We see that the great circle passes less than $3\arcdeg$ from
Hermus, and that it is fairly well aligned with a significant portion
of the stream. 

For the purposes of demonstrating plausibility, we generate orbits
using the Galactic model of \citet{allen1991}. While assuming a
spherical bulge and halo, this model also includes a disk and is
therefore somewhat aspherical at low $|Z|$. As
noted by \citet{eyre2011}, stellar streams are not expected to follow
single orbits in realistic potentials. On the other hand, the
deviations for cold, weakly stripped globular cluster streams in the
nearly spherical potential of the inner halo \citep{kupper2015} are
not expected to be large.

In Panel b of Figure 1 we show a least-squares fit of an orbit to the
RA, dec, and distances of the northern $30\arcdeg$ of Hermus. For
constraining the orbit, we do not use the southern $20\arcdeg$, which
departs both from the great circle of B15, and from the curvature
shown by the northern $30\arcdeg$. G15 noted the different character
of the southern $20\arcdeg$ of Hermus, particularly the change in
curvature of this portion of the stream. While not a definitive test,
we note that an orbit fit to the entire $50\arcdeg$ of Hermus yields a
reduced $\chi^2$ several times higher than a similar fit to just the
northern $30\arcdeg$, with systematic departures in position and
distance at the southern end. We therefore admit the possibility that
the southern $20\arcdeg$ of Hermus is an unassociated structure at the
same distance, and that G14 mistakenly assumed it to be part of the
same structure. This would be in keeping with a common and perhaps
natural tendency to ascribe the complexity of many small features to a
comparatively simple, single stream.

As in G14, we assume uncertainties of $0.3\arcdeg$ in the ten R.A.,
dec positions along the stream, and 3 kpc uncertainties in the
distances at each point. This orbit fit is shown as the red curves in
Figures 1 and 2. Nodal precession of the orbit due to the Galactic
disk naturally brings the Hermus orbit into the plane occupied by
Phoenix.  The orbit fit misses Phoenix by several kpc, but this is
entirely attributable to the distance gradient assumed for
Hermus. Assuming a uniform distance of 20 kpc for every point in
Hermus, we find that a Hermus-only fit lies only 1-2 kpc from the Phoenix
stream.  Reexamining G14's filtered surface density map, we find that
his 15 kpc estimate at the northern end of Hermus relies on some very
faint structures that may or may not be part of the stream. A uniform
20 kpc distance along the northern $30\arcdeg$ appears almost equally
consistent with the data.

Encouraged by these apparent planar alignments, we now include 10
positions along the Phoenix stream in the orbit fit, measured from
Figure 3 of B15. We use 10 positions to give equal weight to both
Hermus and Phoenix. Just as for Hermus, we adopt $0.3\arcdeg$
uncertainties for the positions along the stream, and 3 kpc
uncertainties in the distances at each point. The latter would combine
to give the 0.9 kpc uncertainty claimed by B15 for the entire
stream. We also adopt the distance gradient found by B15, with Phoenix
being $\approx 1$ kpc closer at its northern end than at the southern
end.

This simultaneous fit to the Hermus and Phoenix streams is shown as
the blue curves in Panel b of Figure 1 and in Figure 2. We see that a
single orbit is capable of closely fitting both streams. While the fit
is nearly perfect for Phoenix (including both sky position and
distance gradient), the trajectory in Figure 1 is slightly less curved
than the Hermus stream itself. The maximum deviation between the model
and the stream is $\approx 1.5\arcdeg$ at the extreme northern
end. Given the simplicity of our Galactic model, we do not consider
this very significant. It may be that a slightly prolate or
substructured halo could easily accommodate the curvature of Hermus.

Similar attempts to simultaneously fit the Hyllus and Phoenix streams
are much less interesting. Though Hyllus lies only $4\arcdeg$ east of
Hermus and appears reasonably well aligned with B15's great
circle, the minimum reduced $\chi^2$ is four times larger
than for Hermus. Orbital precession is evidently insufficient to match
the trajectories of both streams simultaneously, and the best-fitting
orbits are clearly out-of-plane for one or the other, typically
missing by several kpc.

The orbit that simultaneously fits Hermus and Phoenix is fairly
circular, with apo- and perigalactica of $19.3^{+1.7}_{-0.4}$ and
$17.6^{+0.2}_{-0.8}$ kpc and an eccentricity of only $\approx
0.05$. This is somewhat surprising, given the roughly isotropic
orbital distribution seen among globular clusters. Moreover, it argues
that the progenitor of a putative Phoenix-Hermus stream must have been
rather loosely bound to have been so significantly depleted in such a
relatively benign orbit.  The orbit is inclined $\approx 60\arcdeg$ to
the Galactic plane, and the Phoenix stream is situated very nearly at
the apogalactic point of the orbit. This would be qualitatively
consistent with the apparent fading out of the Phoenix stream at its
northern end, where the stars are picking up speed and the stream
consequently becomes more tenuous.

B15 found two overdensities in Phoenix that are slightly out of
alignment with the bulk of the stream and suggest that these
overdensities could be the remnants of the progenitor. If true, then
the arrangement of these overdensities implies that the stream must be
moving from south to north and is in a prograde orbit around the
Galaxy. The orbit that best fits both streams passes south from
Hermus, through the disk on the far side of the Galactic center,
within $15\arcdeg$ of the south celestial pole, and then north to
Phoenix. If Phoenix and Hermus are related, then Hermus would
constitute the trailing tail of the Phoenix stream. From the northern
end of the leading (north) arm of the Phoenix stream to the trailing
(northern) end of the Hermus stream, the best-fit orbit subtends
$183\arcdeg$ on the sky and 76 kpc through space. Viewed from the  
Galactic center, the orbit would subtend $235\arcdeg$, or two thirds
of a complete wrap around the Galaxy.

The leading arm passes north from Phoenix, through the anticenter
portion of the disk, within $10\arcdeg$ of the north celestial pole,
and then south to Hermus. However, nodal precession causes its path to
deviate $\approx 9\arcdeg$ from the orbital plane of the stream. This
is shown in Figure 3, where we show a nearly edge-on view of the
orbit.

Owing to the near-circularity of the best-fit orbit, the leading arm
ends up at nearly the same distance and nearly parallel to Hermus in
the SDSS footprint. It lies $10\arcdeg$ west of Hermus on the northern
end (right hand side) of Figure 1 but converges with Hermus near Pal
5. We have carefully examined this region, and while there are some
faint features that roughly line up with the predicted orbit of the
leading arm, we find nothing we would have identified as a stream at
the outset.  This could be an indication that the leading and trailing
arms are of different lengths, that the leading arm is
weaker than the trailing arm, or that the leading arm has been
dispersed by an encounter with one or more massive substructures. Of
course, it could also be an indication that Hermus and Phoenix are not
physically associated.

We have compared our best-fitting orbit with the positions of all
known Galactic globular clusters. The closest matches are with Pal 1, NGC
1261, and Pal 5. NGC 1261 and Pal 5 are clearly ruled out by B15 and
Figure 1 of this work. Similarly, Pal 1 lies $\approx 2.3$ kpc
laterally from the nearest branch of the orbit. Owing to Pal 1's
proximity to the Sun, the best-fitting orbit passes no closer than
$12\arcdeg$ from the cluster. We conclude that none of the known
globular clusters is likely to be the progenitor of this stream.

\section{Predicted Observables} \label{predictions}

If Hermus and Phoenix are indeed part of the same stream, then our
orbit model can be used to make approximate predictions for the radial
and tangential velocities we would expect to measure. Figure 4 shows
positions, distances, radial velocities, and proper motions as a
function of Galactic longitude for both prograde and retrograde
orbits. We show only the portion of the orbit connecting the streams on
the far side of the Galactic center, which is the arm that, via
nodal precession, correctly predicts the orientations of both Hermus and
Phoenix.

Uncertainties were estimated using the marginal $\chi^2$ distributions
for $v_{hel}, \mu_{\alpha} cos(\theta)$, and $\mu_{\delta}$. Each
parameter was offset to its 90\% confidence limit and the other fit
parameters were then varied to find a new $\chi^2$ minimum. The shaded
regions in Figure 4 encompass the entire range of observable
parameters resulting from this procedure.

Also shown in Figure 4 are Galactic standard of rest velocity cuts
used by Martin et al. (2016) to kinematically identify Hermus using
blue horizontal branch stars. The velocities clearly favor a prograde
orbit for Hermus and, if the Phoenix and Hermus streams
are part of the same stream, are completely consistent with our
predictions for a prograde orbit.

\section{Conclusions} \label{conclusions}

Having found that the Hermus and Phoenix streams are nearly coplanar,
we have investigated whether a single orbit could accommodate both
streams.  Using a Galactic model with a disk, bulge, and spherical
halo, we find that we can indeed match the trajectories of both
streams with a single orbit. Moreover, this match is partly
facilitated by orbital precession, which naturally brings the orbital
plane into alignment with each stream within half an orbit around the
Galaxy.

B15 identify a possible progenitor within the Phoenix stream. The
misalignment of this feature with the stream itself suggests that
Phoenix is on a prograde orbit around the Galaxy. If Phoenix and
Hermus are part of the same stream, then Hermus must be a remote part
of the trailing arm.

While not proven, we consider the hypothesis that Hermus and Phoenix
are part of the same stream as entirely plausible. Confirmation will
require spectroscopy of stars in both Hermus and Phoenix. The
kinematic discovery of Hermus \citep{martin2016} already supports a
prograde orbit for the stream, and is entirely consistent with our
predictions based on a single-stream orbit.

If radial velocity measurements and/or proper motions confirm a
physical association between the Hermus and Phoenix streams, then as
the longest cold stream yet discovered, Phoenix-Hermus would provide us
with a remarkable new probe of the Galactic potential.

\acknowledgments

We thank an anonymous referee for several useful suggestions that
improved both the clarity and thoroughness of the manuscript.

\clearpage

\begin{figure}
\epsscale{1.0}
\plotone{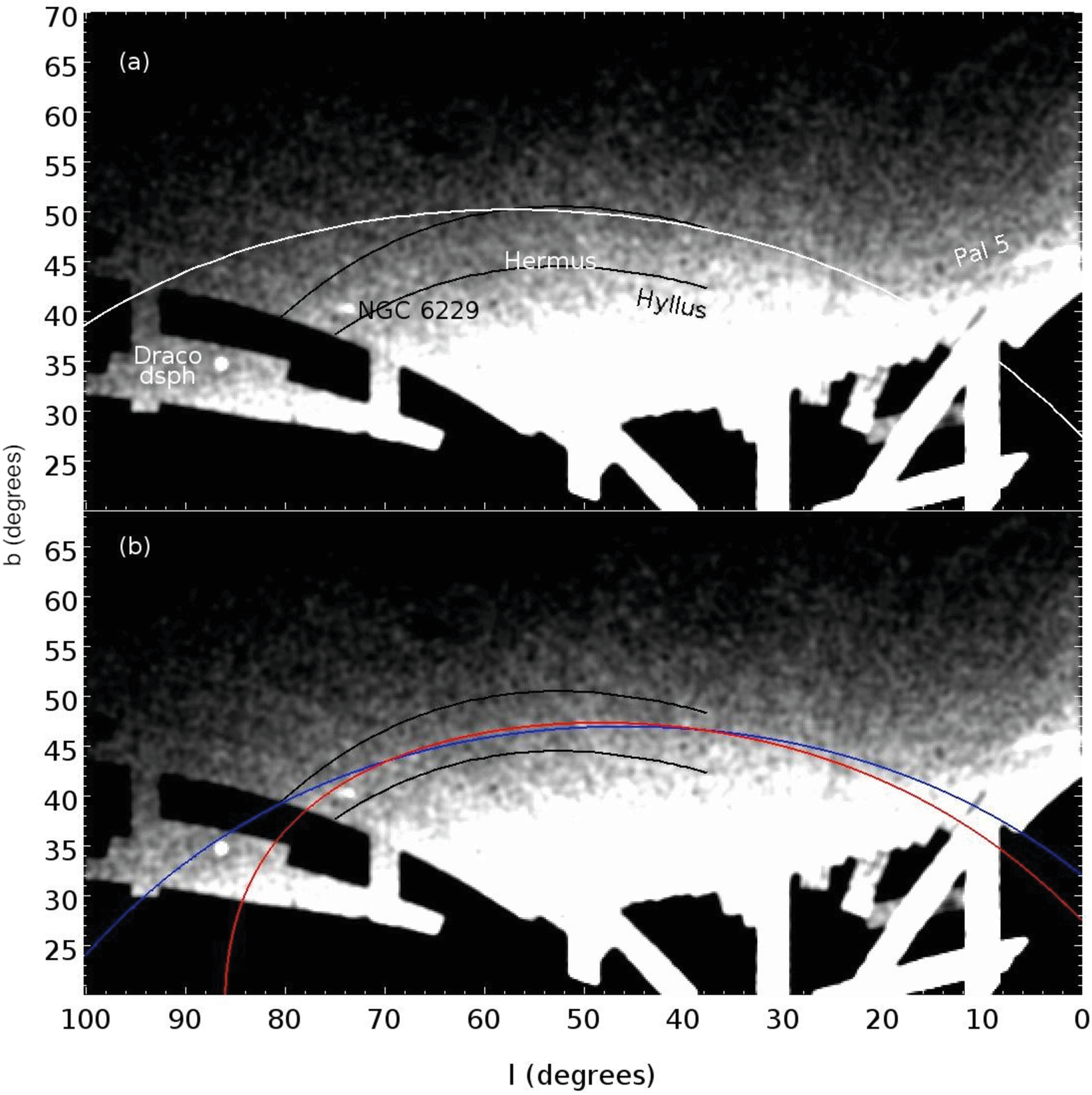}
\caption{Panel a: filtered surface density map of the western portion
  of the northern SDSS footprint, in Galactic coordinates.  The
  stretch is linear, with lighter areas indicating higher surface
  densities. The map is the result of a filter based on the
  color-magnitude distribution of stars in the globular cluster M 53,
  and shifted to a distance of 20 kpc. The map has been smoothed with
  a Gaussian kernel of $0.6\arcdeg$. The black curves correspond to
  the Hermus trajectory of G14, offset $\pm 3\arcdeg$ in {\it b}. The white curve
  is the Galactocentric great circle fit to Phoenix by B15. Panel b:
  The same map with best-fit orbits shown. The red curve shows an orbit fit
  to only the northern $30\arcdeg$ of Hermus, while the blue curve
  shows a simultaneous fit to both the Hermus and Phoenix streams. }

\end{figure}

\begin{figure}
\epsscale{2.0}
\plotone{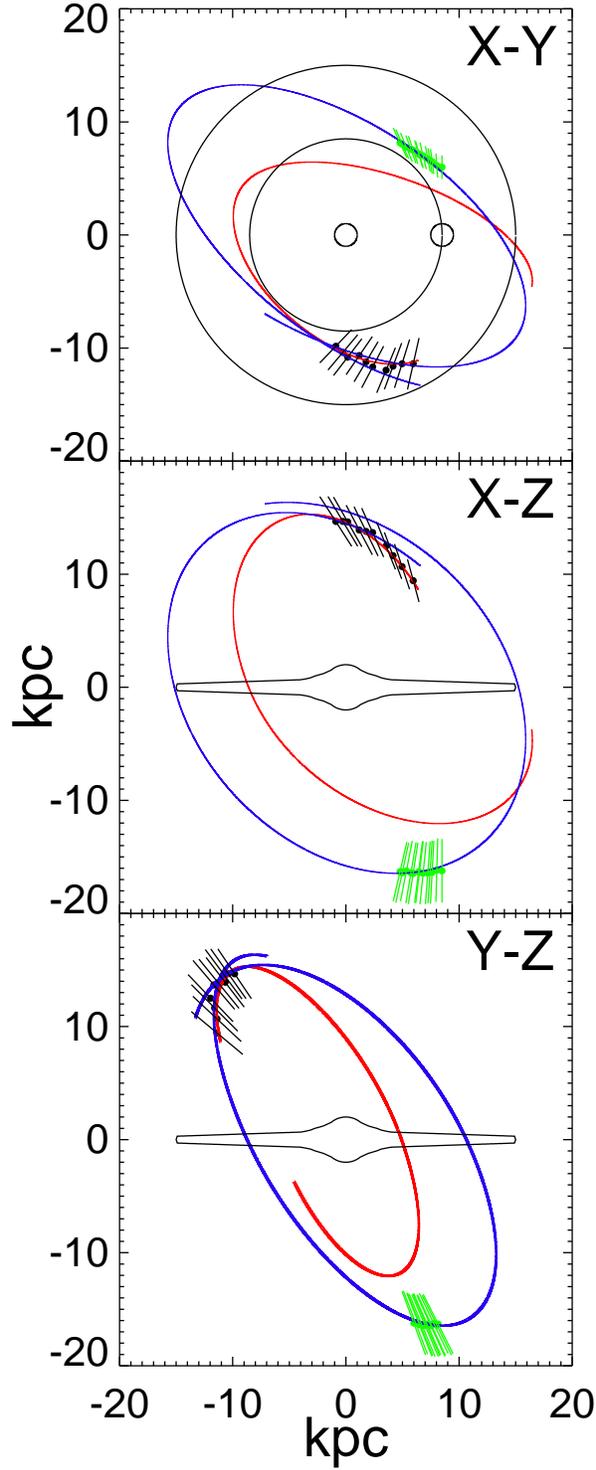}
\caption{Projections of best-fit orbits in Galactic cartesian
  coordinates. The solar circle and the position of the sun are
  indicated. The path of the Hermus stream is shown by the black points,
  where the error bars correspond to $\pm 3$ kpc. The green points and
  error bars indicate the path of the Phoenix stream. The red curve is
  an orbit fit solely to Hermus, while the blue curve shows a
  simultaneous fit to both Hermus and Phoenix. Note that orbital precession
  due to the disk is required to bring the orbit into alignment with
  the Phoenix stream.}

\end{figure}

\begin{figure}
\epsscale{1.0}
\plotone{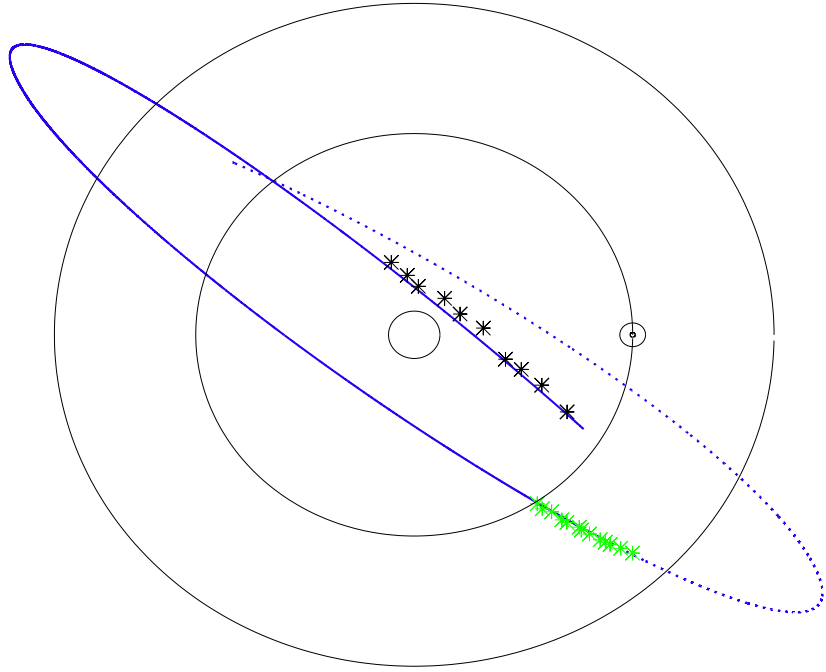}
\caption{A projection of the best-fit orbit viewed nearly in the
  orbital plane. The leading (dotted) and trailing (solid) tails are
  at nearly the same radius, but are inclined by $\approx 9\arcdeg$
  to one another due to nodal precession. Hermus is shown by black asterisks
  while Phoenix is shown in green. Scale and orientation are provided
  by the solar circle, with the Sun at x = 8.5 kpc.}

\end{figure}

\begin{figure}
\epsscale{1.0}
\plotone{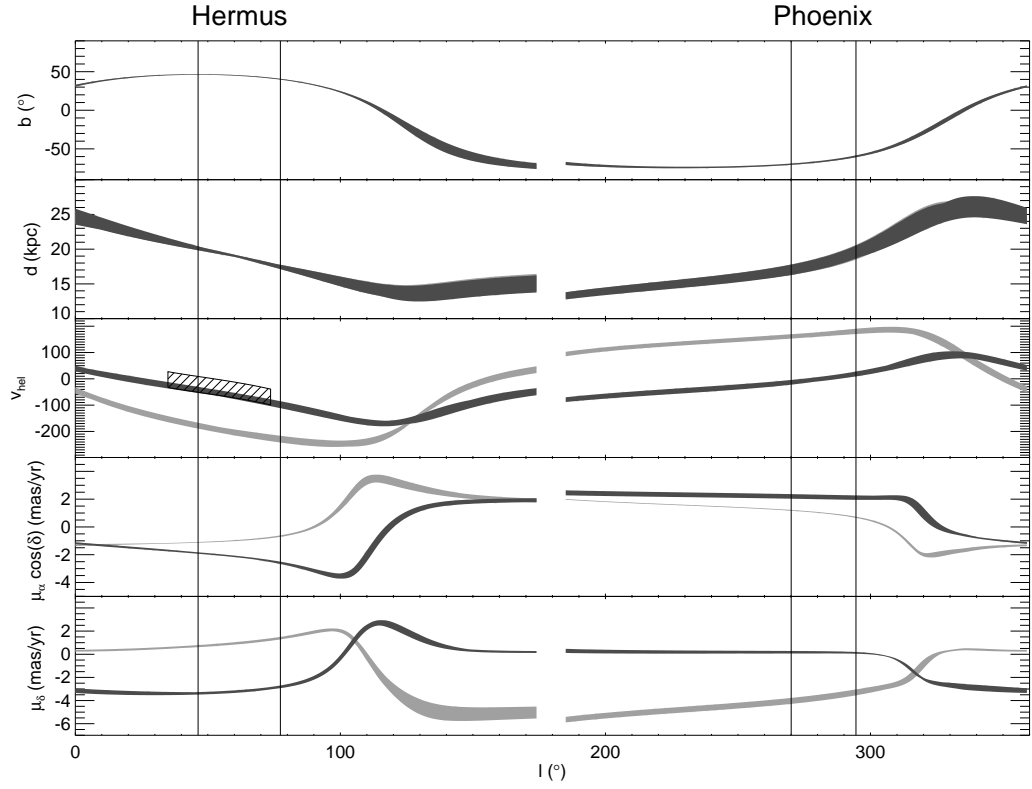}
\caption{Predictions of our model orbit for various observables, as a
  function of Galactic longitude. The shaded regions encompass the
  90\% confidence interval in each case. The dark gray shaded regions
  correspond to a prograde orbit, while the lighter gray regions
  indicate expectation values for a retrograde orbit. The endpoints of
  the Hermus and Phoenix streams are shown by the vertical lines. The
  hashed box in the middle panel shows the $v_{GSR}$ limits used by
  Martin et al. (2016) in kinematically detecting Hermus with blue
  horizontal branch stars. In the case of a prograde orbit, both
  streams would be moving towards lower longitudes, while in the retrograde
  case they would be moving to the right towards higher l.}

\end{figure}


\begin{thebibliography}{}

\bibitem[Ahn(2014)]{ahn2014} Ahn, C. P., Alexandroff, R., \& Allende Prieto, C. et al. 2014, \apjs, 211, 17

\bibitem[Allen \& Santillan(1991)]{allen1991} Allen, C., \& Santillan,
A. 1991, {\it Rev. Mex. Astron. Astrofis.}, 22, 255

\bibitem[Balbinot et al.(2015)]{balbinot2015} Balbinot, E., Yanny, B.,
  Li, T. S. et al. 2015, (B15) \apj, in press.

\bibitem[Carlberg(2009)]{carlberg2009} Carlberg, R. G., \apj, 705, 223

\bibitem[Carlberg \& Grillmair(2013)]{carlberg2013} Carlberg, R. G.,
  \& Grillmair, C. J. 2013, \apj, 768, 171

\bibitem[Eyre \& Binney(2010)]{eyre2011} Eyre, A., \& Binney, J. 2011,
  \mnras, 413, 1852

\bibitem[Fardal et al.(2015)]{fardal2015} Fardal, M. A., Huang, S., \&
  Weinberg, M. D. 2015, \mnras, 452, 301

\bibitem[Grillmair \& Dionatos(2006)]{grillmair2006} Grillmair, C. J., \&
Dionatos, O. 2006, \apjl, 643, 17

\bibitem[Grillmair \& Carlin(2016)]{grillmair2016} Grillmair, C. J.,
  \& Carlin, J. L. 2016, in {\it Tidal Streams in the Local Group and
    Beyond}, H. J. Newberg \& J. L. Carlin eds., Springer

\bibitem[Koposov et al.(2010)]{koposov2010} Koposov, S., Rix, H-W, \&
  Hogg, D. W. 2010, \apj, 712, 260

\bibitem[K{\"u}pper et al.(2012)]{kupper2012} K{\"u}pper, A. H. W., Lane,
  R. R., \& Heggie, D. C. 2012, \mnras, 420, 2700

\bibitem[K{\"u}pper et al.(2015)]{kupper2015} K{\"u}pper, A. H. W.,
  Balbinot, E., Bonaca, A., Johnston, K. V., Hogg, D. W., Kroupa, P.,
  \& Santiago, B. X. 2015, \apj, 803, 80

\bibitem[Lane et al.(2012)]{lane2012} Lane, R. R., K{\"u}pper, A. H. W.,
  \& Heggie, D. C. 2012, \mnras, 423, 2845

\bibitem[Martin et al.(2016)]{martin2016} Martin, C., et al. 2016, in prep.

\bibitem[Odenkirchen et al.(2003)]{odenkirchen2003} Odenkirchen,
  M. et al. 2003, \aj, 126, 2385

\bibitem[Rockosi et al.(2002)]{rock2002} Rockosi, C. M. et al. 2002,
\aj, 124, 349

\bibitem[Smith(2016)]{smith2016} Smith, M. C. 2016, in {\it Tidal
  Streams in the Local Group and Beyond}, H. J. Newberg \&
  J. L. Carlin eds., Springer

\bibitem[Willett et al.(2009)]{willett2009} Willet, B. A., Newberg,
  H. J., Zhang, H., Yanny, B., \& Beers, T. C. 2009, \apj, 697, 207

\bibitem[Wyse(2009)]{wyse2009} Wyse, R. F. G 2009, IAU Symp. 258, 11

\bibitem[Yoon et al.(2011)]{yoon2011} Yoon, J. H., Johnston, K. V., \&
  Hogg, D. W. 2011, \apj, 731, 58

\end{thebibliography}
\end{document}